\documentclass[namedreferences]{solarphysics}
\usepackage[optionalrh,solaromanenum]{spr-sola-addons} 
\usepackage{graphicx}                    
\usepackage{color}                       
\usepackage{url}                         

\begin{document}

\begin{article}

\begin{opening}

\title{Cycle 23 Variation in Solar Flare Productivity} 

\author{Hugh~\surname{Hudson}$^{1,2}$\sep
Lyndsay~\surname{Fletcher}$^2$\sep
Jim~\surname{McTiernan}$^1$
}

%

\institute{$^1$ SSL, University of California, Berkeley, CA, USA;
                     email: \url{hhudson@ssl.berkeley.edu} \\ 
             $^2$ School of Physics and Astronomy, University of Glasgow, UK\\}

%

\runningauthor{Hudson {\it et al.}}
\runningtitle{Flare productivity statistics}

\begin{abstract}
The NOAA listings of solar flares in cycles 21-24, including the GOES soft X-ray magnitudes, enable a simple determination of the number of flares each flaring active region produces over its lifetime.
We have studied this measure of flare productivity over the interval 1975--2012.
The annual averages of flare productivity remained approximately constant during cycles 21 and 22, at about two reported M or X flares per region, but then increased significantly in  the declining phase of cycle 23 (the years 2004-2005).
We have confirmed this by using the independent RHESSI flare catalog to check the NOAA events listings where possible.
We note that this measure of solar activity does not correlate with the  solar cycle.
The anomalous peak in flare productivity immediately preceded the long solar minimum between cycles 23 and 24.
\end{abstract}


\end{opening}

\section{Introduction}  
The unusual behavior of solar activity during the sunspot minimum of 2008 has excited much interest.
In general solar activity does exhibit long-term variability, extending to time scales exceeding that of the Hale cycle.
Albregtsen and Maltby (1981) found that the ratio of umbra/photosphere brightness ratio varied with phase in the solar cycle; recently
Penn and Livingston (2006) have noted a systematic change in umbral magnetic field intensity as well (\textit{cf.} Watson \textit{et al.}, 2011).
\nocite{2006ApJ...649L..45P,1981SoPh...71..269A,2011A&A...533A..14W}
The latter discovery spans the maximum of cycle 23, leading up to the unexpectedly extended minimum between cycles 23 and 24 (see, {\it e.g.}, papers in IAU Symposium No. 286; Mandrini and Webb 2012).
\nocite{2012IAUS..286.....M}
This ``extended minimum'' had precedents, but not within the modern era (roughly speaking, beginning with the introduction of the F10.7 index by Covington in 1947; see, \textit{e.g.}, Tapping (1987).
\nocite{1987JGR....92..829T}
Accordingly much new information has surfaced, ranging from variations away from a supposed basal level of total solar irradiance in the minima \cite{2011SSRv..tmp..133F} to an unprecedented level of cosmic-ray flux \cite{2010ApJ...723L...1M}.
The various observations suggest the existence of heretofore unknown properties of the solar magnetic field and its variation,
both on global and local scales.

In this paper we report another effect: the variation of the flare productivity of a given active region.
Here we use this term simply to mean the number of flares per active region
(\textit{cf.} Abramenko, 2005), rather than anything to do with a region's magnetic structure.
There is an extensive literature identifying flare occurrence with local properties of an active region, including both intrinsic properties such as helicity injection during flux emergence ({\it e.g.}, Heyvaerts and Priest, 1984; Rust, 1994; Low, 1996) and global properties such as the coronal environment of an active region ({\it e.g.}, T{\" o}r{\" o}k and Kliem, 2005; Dalla \textit{et al.}, 2007; Jing \textit{\textit{et al.}} 2010).
\nocite{2005ApJ...629.1141A,2010ApJ...713..440J}
\nocite{1984A&A...137...63H,1994GeoRL..21..241R,1996SoPh..167..217L}
\nocite{2005ApJ...630L..97T}
\nocite{2007A&A...468.1103D}
Without prejudging the observational evidence for such processes, we have simply studied the NOAA and RHESSI ({\it Reuven Ramaty Solar Spectroscopic Imager}) databases on flare occurrence as seen in soft X-rays by the GOES ({\it Geostationary Operational Environmental Satellite}) detectors;
see Wheatland (2001) for background information on this approach.
\nocite{2001SoPh..203...87W} 
The number of major flares per active region remained approximately constant during cycles 21 and 22, but then exhibited a distinct variation towards the end of cycle~23 as described here.
For the initial period of cycle~24, to the time of writing at the end of 2012, the flare productivity appears to have returned to its prior levels.

\section{NOAA Database}
\begin{figure}                                                   
\centerline{\includegraphics[width=0.8\linewidth] {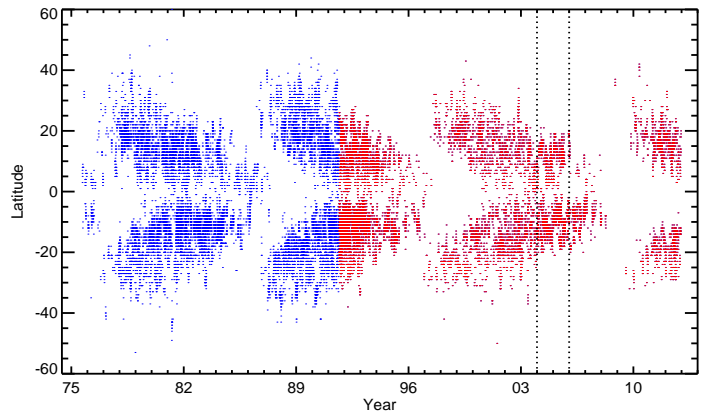}}
\caption{The full NOAA database, as extracted from the pre-SolarSoft files (blue) and the SolarSoft database (red), the latter taken as definitive.
The vertical lines mark the years 2004-2005.
}    
\label{fig:gev_all_1}
\end{figure}

The primary source of information for our assessment of flare productivity is the NOAA ``events'' database.
Since 1991 a version of these data has been directly available via SolarSoft \cite{1998SoPh..182..497F}; the earlier databases used here began in 1975 and were obtained directly from NOAA Web archives.
We have adopted the GOES soft X-ray classification as a standard referenceand accumulate statistics separately for the C and (M,X) ranges.
Note that the soft X-ray photometers on the GOES spacecraft have differed slightly from one to another over the years, but that cross-calibration has generally been possible ({\it e.g.}, White {\it et al.}, 2005).
We return to the issue of database reliability in Section~\ref{sec:rhessi}.
\nocite{2005SoPh..227..231W}
Figure~\ref{fig:gev_all_1} shows the total content of these databases, plotting all C, M, and X-class flare positions vs time.
These records extend from 1~September 1975, through 11~December 2012, and (redundantly) contain 82,344 total entries of all GOES classes. 
We screened these by eliminating redundancy and be taking only events for which the NOAA region number was listed, and also by removing  some obvious outliers in heliolatitude.
This step was purely cosmetic, since all of the 86 flares thus eliminated were B-class events and our
work here is with C-class and above.
Finally we eliminated events with heliolongitude outside the range $\pm75^\circ$,  in order
to minimize the effects of solar tilt at the extreme limb.
This resulted in a sample of (20,143,  3,791, and 345) flares of (C, M, and X)-class respectively.

\begin{figure}                                                   
\includegraphics[width=0.49\linewidth] {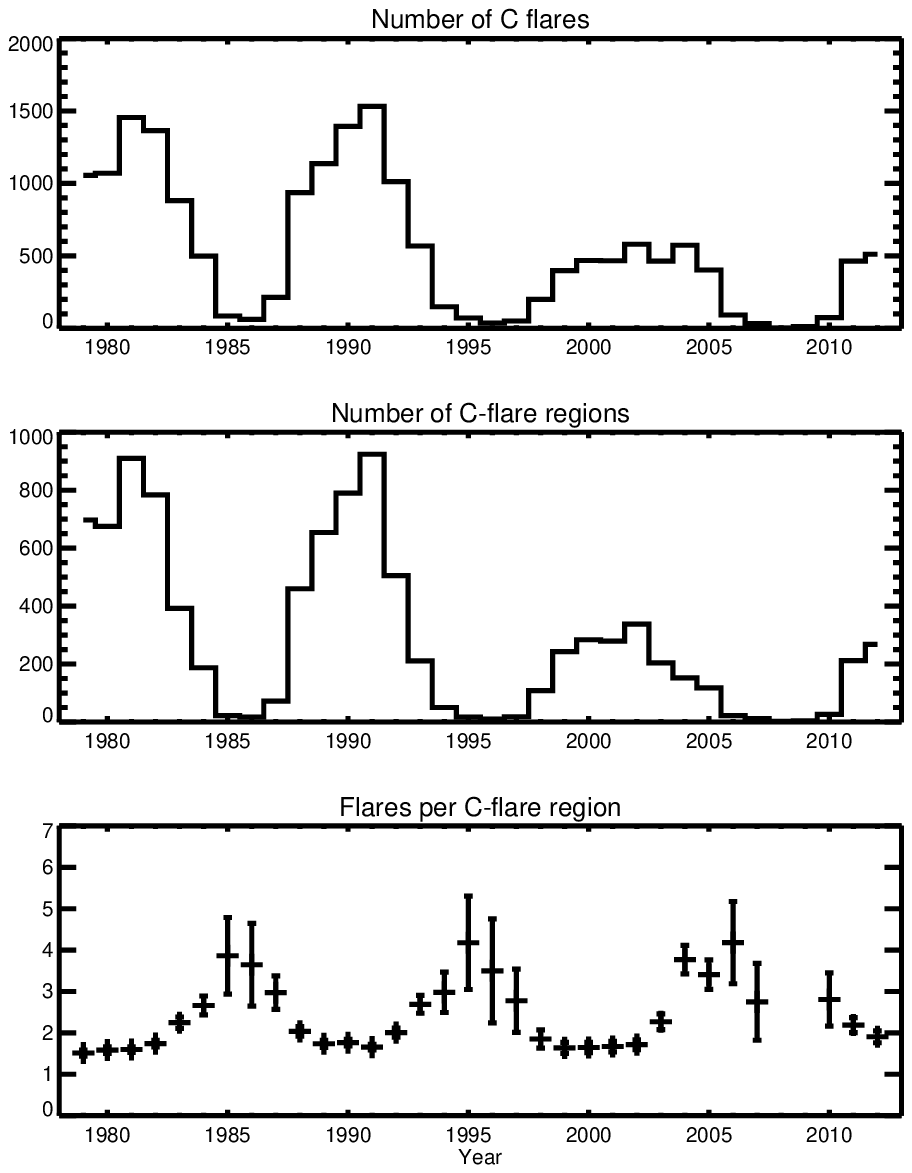}
\includegraphics[width=0.49\linewidth] {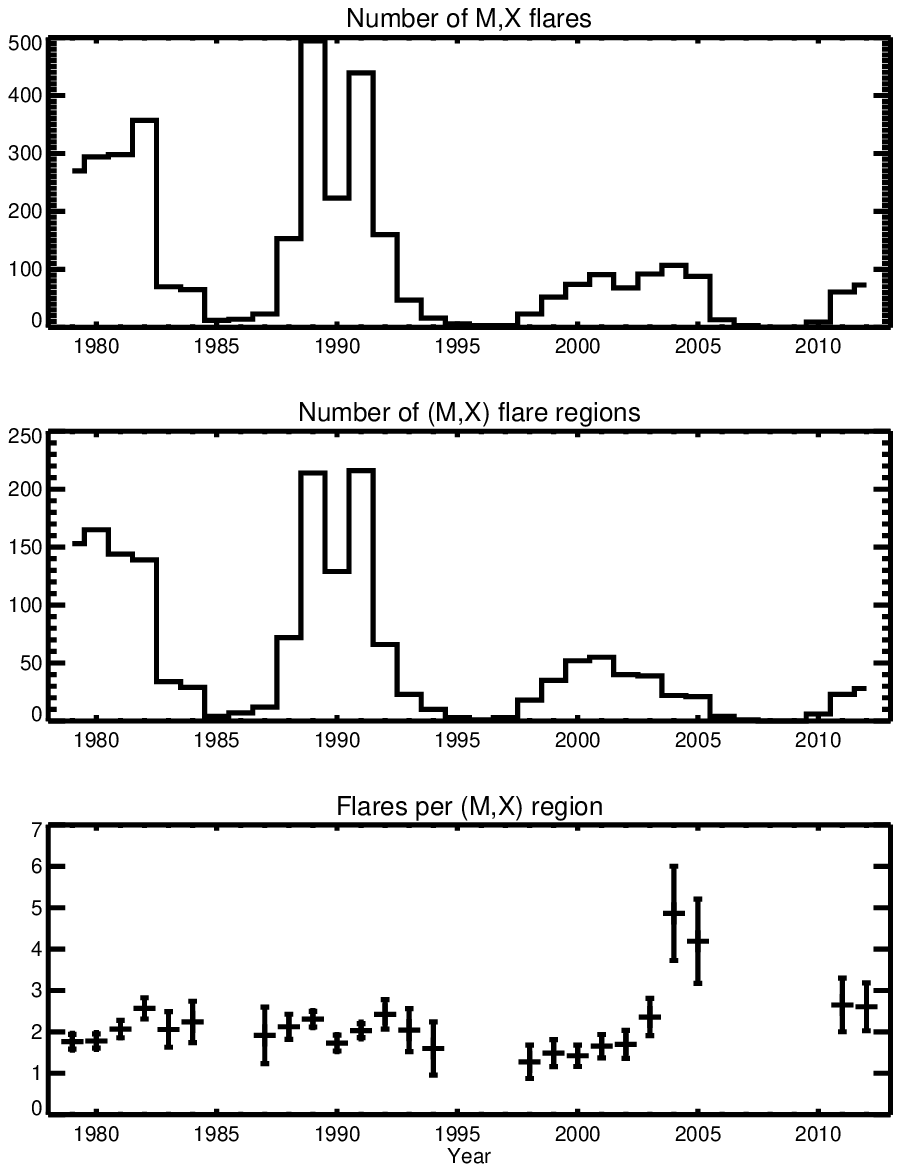}
\caption{Flare productivity for individual active regions, shown on the left for C-class events, and on the right for M and X.
The striking systematic variation in the lower-left panel is an artifact due to the obscuration of weaker flares by stronger ones (Wheatland, 2001).
Years with fewer than ten regions have been omitted for clarity here.
}    
\label{fig:gevall}
\end{figure}

We have looked at the flare productivity by region by simply plotting the number of flares identified with a given flaring region as one-year averages (Figure~\ref{fig:gevall}).
We plot C and (M,X)-classes separately because of the systematic undercounting at C-class: the higher background produced by a major flare makes it harder to detect a concurrent minor one, or one that occurs during the gradual decay of such an event.
Wheatland (2001) discusses this effect in detail and terms it ``obscuration.''
Our simple distinction does not eliminate the bias completely because it depends on the flare occurrence pattern, which is definitely not random. 
However the arbitrary break point between GOES C- and M-class flares allows a qualitative distinction in the solar-cycle pattern of flare productivity to emerge.

Figure~\ref{fig:gevall} shows that the region productivity of (M,X)-class flares remained near or below two flares per region, but with a significant increase in 2004-2005.
Similar increases did not occur at the corresponding phases of cycles~21 or~22.
This anomalous behavior is formally significant at above the 3$\sigma$ level.
In the first two annual means of cycle~24, the flare productivity returned to earlier levels.

\begin{figure}                                                   
\includegraphics[width=0.49\linewidth] {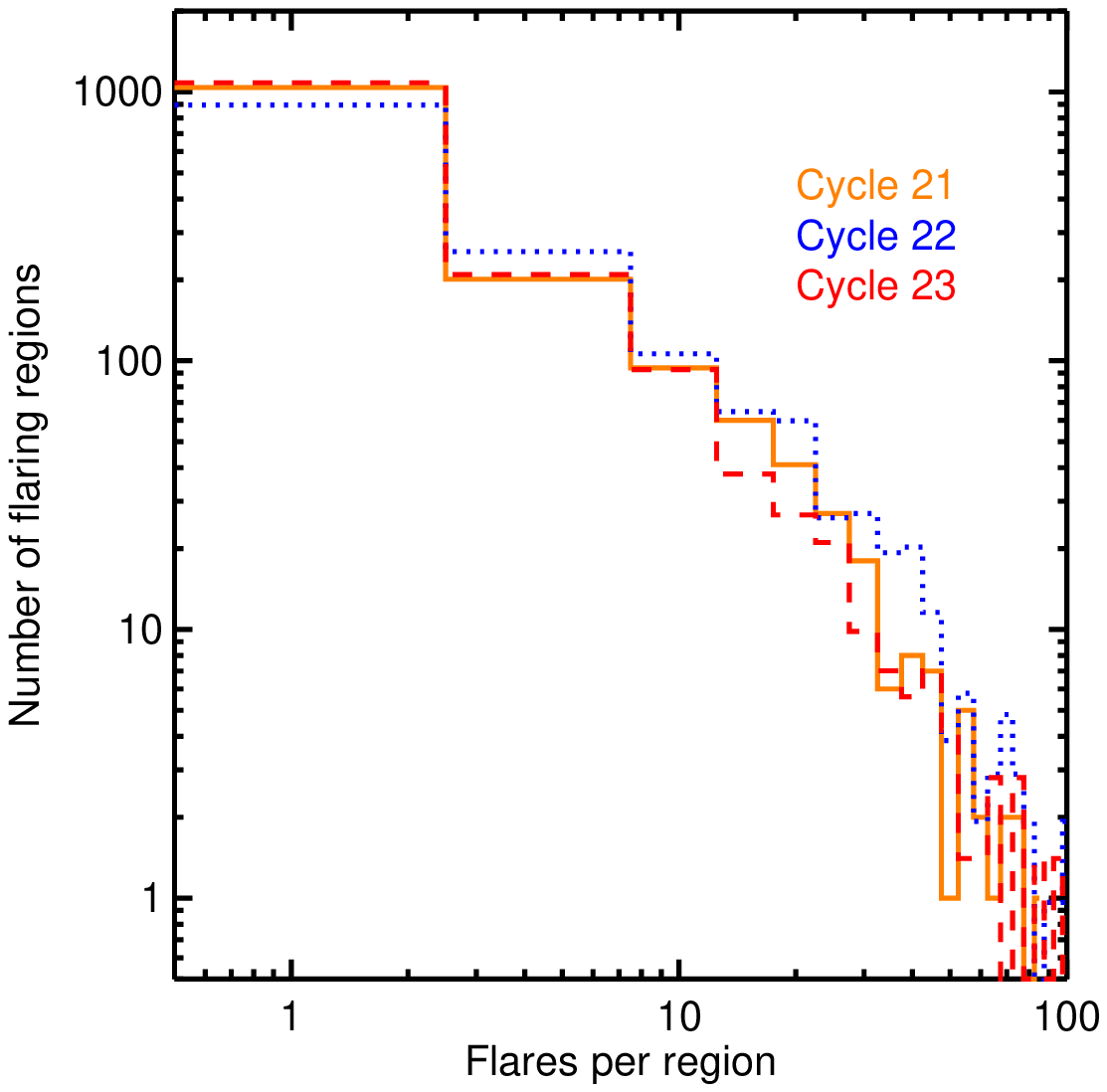}
\includegraphics[width=0.49\linewidth] {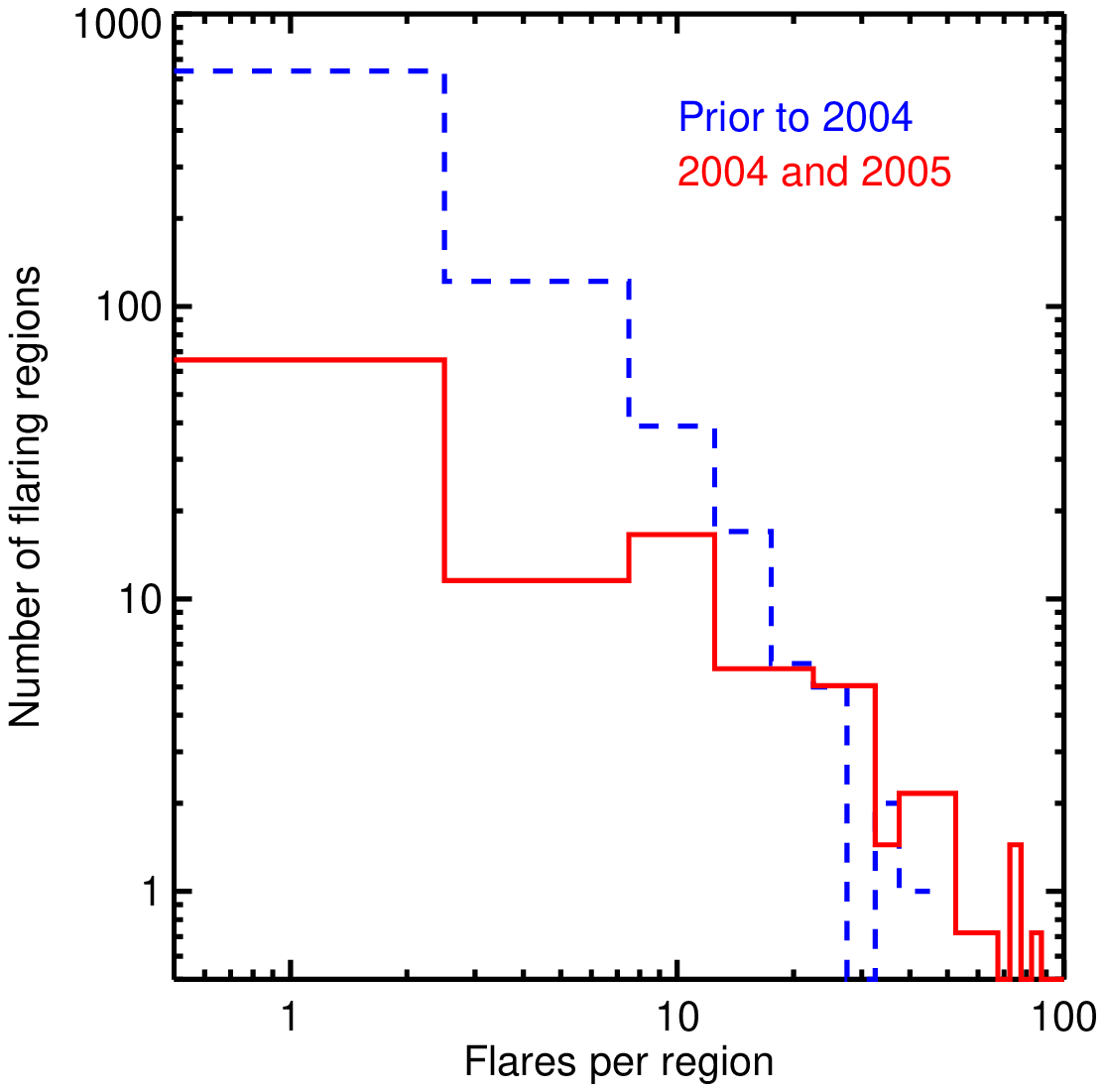}
\caption{Histograms of flare productivity, comparing cycles 21, 22, and 23 (left; the orange solid line is cycle 21, the blue dotted line cycle 22, and the red dashed line cycle 23), and two epochs in  cycle 23 (right; blue dashed line for data prior to 2004, red solid line for 2004-2005).
The histograms have been equalized by normalizing to the number of events in cycle 21 on the left, and to the number of events in the epoch prior to 2004, in cycle 23,  on the right.
}    
\label{fig:histograms}
\end{figure}

Figure~\ref{fig:histograms} give a different view of the flare productivity by region, in the form of histograms of productivity values for individual regions.
These amplify the results found in Figure~\ref{fig:gevall} by showing the distribution in flare productivity of the most productive regions.
They reveal a strong flattening of the distribution of flare productivity in 2004-2005, which confirms the conclusion drawn from the mean flare productivities.

\begin{figure}  
\centering                                                 
\includegraphics[width=0.7\linewidth] {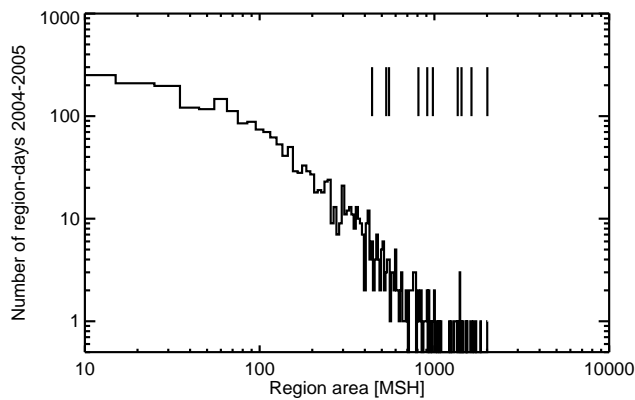}
\caption{Histogram of region areas (2004-2005), with vertical lines showing the maximum areas of the most flare-productive regions during the latter half of cycle 23.
}    
\label{fig:region_hist}
\end{figure}

Table~1 lists the ten most flare-productive regions in the second half of the cycle~23 maximum, with peak sunspot areas and magnetic classifications on the date of maximum area.
``MSH'' refers to the peak group sunspot area listed, in millionths of the solar hemisphere,
and the magnetic classification refers to the time of this peak area.
As expected from previous studies of flaring patterns (\textit{e.g.}, Gaizauskas, 1982; Zirin and Liggett, 1987; Schrijver, 2007), these regions were all complex and mostly (8/10) classified as having $\delta$~configurations.
\nocite{1982AdSpR...2...11G}
\nocite{1987SoPh..113..267Z}
\nocite{2007ApJ...655L.117S}
They were not necessarily the regions with the largest areas, though, underscoring the requirement for other factors (\textit{e.g.}, magnetic complexity) in flare productivity.
These regions produced 29 of the 30 X-class flares reported during this interval.
Figure~\ref{fig:region_hist} compares the areas of the flare-productive regions with the histogram of areas for all region reports in the latter half of cycle 23.

\begin{table}\label{tab:odd}
\caption{The most flare-productive active regions in the latter half of cycle 23}
\bigskip
\begin{tabular}{l r r c r r r}
\hline
NOAA & Date & Area [MSH] & Hemisphere & Class & M & X \\
\hline
   10536 & 7 January 2004 & 980 & S &$\beta\gamma\delta$ &1 & 1 \\
   10635 & 20 June 2004 & 550 & S &$\beta\gamma$ &1 & 1\\
   10649&17 July 2004&     530&S&$\beta\gamma\delta$ & 10 & 6 \\
   10652&22 July 2004&    2010&N&$\beta\gamma\delta$& 16 & 1\\
   10656&11 August 2004&    1360&S&$\beta\gamma\delta$ & 20 & 1\\
   10696& 5 November 2004&     910&N&$\beta\gamma\delta$& 11 & 2\\
   10720&15 January 2005&    1630&N&$\beta\delta$& 15 & 5\\
   10786& 8 July 2005 & 420& N & $\beta\gamma\delta$ & 5 & 1 \\
   10808&13 September 2005&    1430&S&$\beta\gamma\delta$ & 19 & 9 \\
   10822 & 18 November 2005 & 810 & S & $\beta\gamma$ & 5 & 1 \\
 \hline
\end{tabular}
\end{table}

\section{RHESSI Observations}\label{sec:rhessi}

\begin{figure}                                                   
\includegraphics[width=0.49\linewidth] {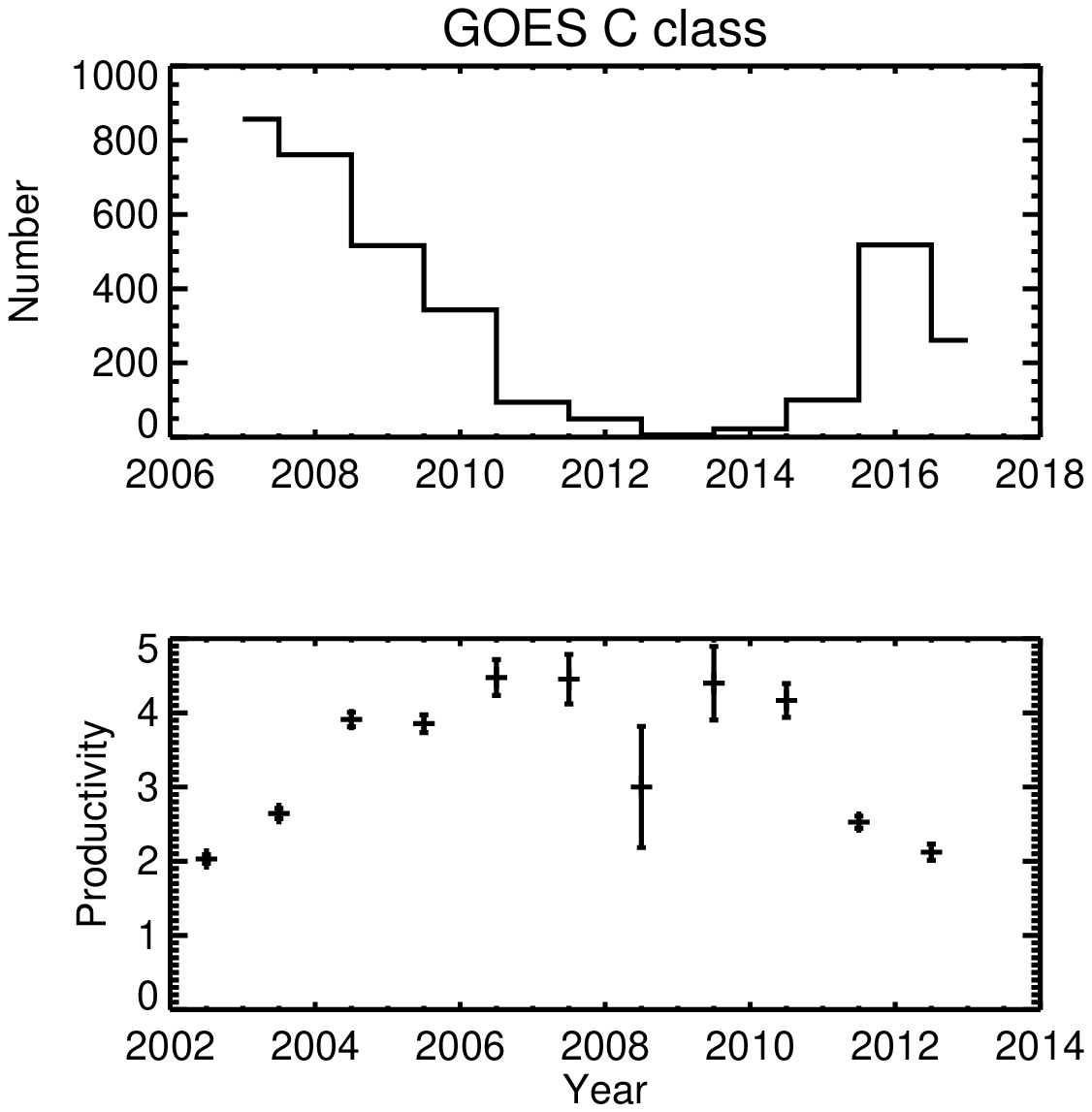}
\includegraphics[width=0.49\linewidth] {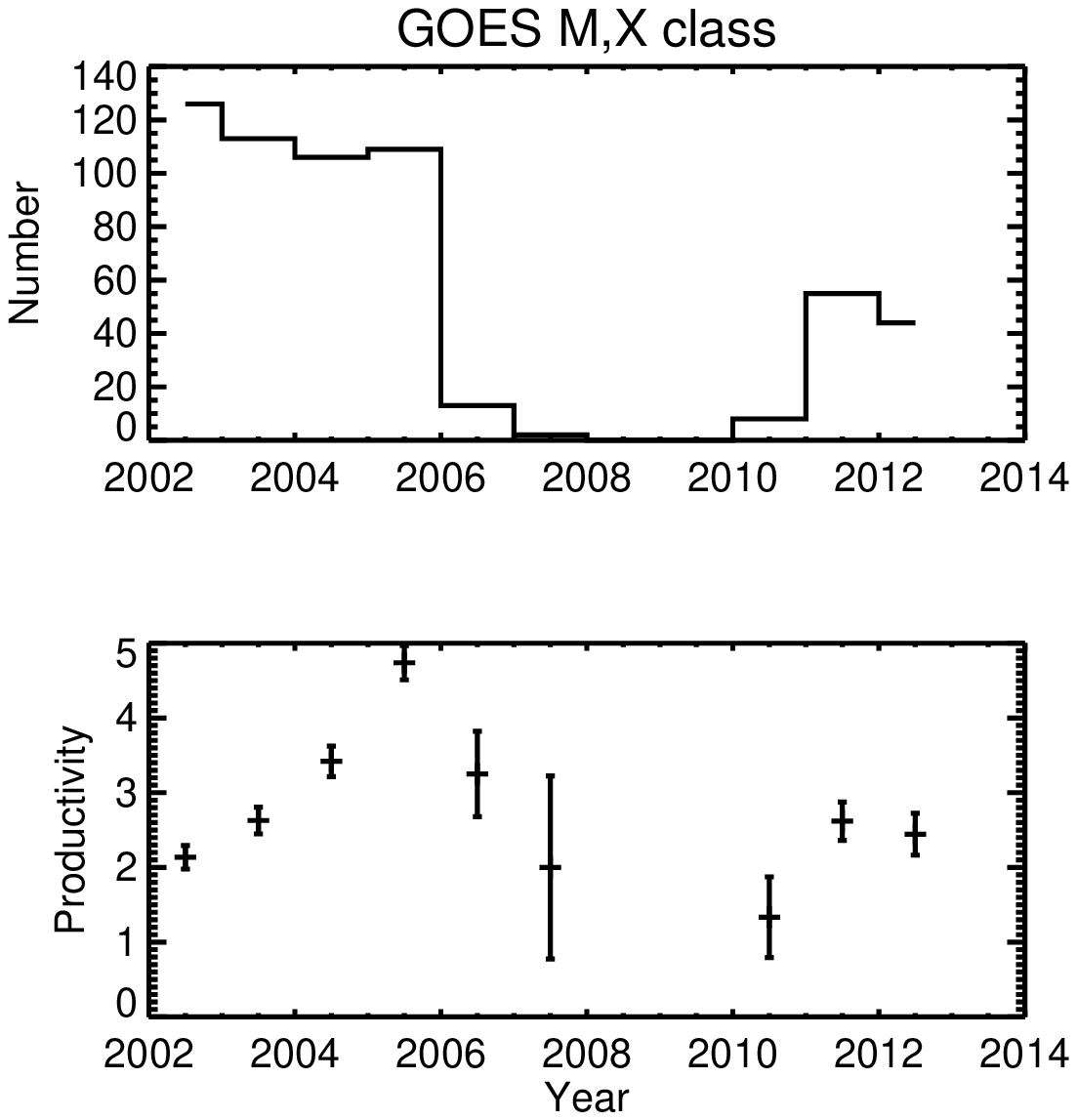}
\caption{Numbers of flares and mean flare productivities as inferred from the RHESSI flare catalog: left, C-class;  right, (M,X)-class.
The lower-right panel of this Figure can be compared with the lower-right panel of Figure~\ref{fig:gevall}.
}    
\label{fig:rhessi}
\end{figure}

The databases we have used have been generated outside our control, and probably contain both random and systematic errors.
The early databases appear to have had some keyboard entry errors as well.
We have therefore sought to check the recent data against the RHESSI flare catalog (available through SolarSoft or as a text file at \url{http://hesperia.gsfc.nasa.gov/hessidata/dbase/hessi_flare_list.txt}).
This provides an independent location of each event, but done entirely through an automated pipeline reduction that identifies RHESSI flare events and tabulates their properties.
The positions are unambiguous and come from direct imaging in higher-energy X-rays than GOES is sensitive to.
The hard X-ray sources are physically smaller and thus better defined for this purpose.
The RHESSI data began in 2002, and so we can check the annual mean flare productivity in the same manner as for the NOAA database.
Note that the original NOAA flare listings derived positions from H$\alpha$ flare associations.
We find, in Figure~\ref{fig:rhessi}, a confirmationn of the NOAA flare-productivity anomaly in cycle~23 by use of this independent database.

We note also that the RHESSI flare identifications show an increase in flare productivity at the C-class level, beginning roughly with the October 2003 events and continuing through the extended solar minimum period.
We do not think that this effect could be seen in the lower-left panel of Figure~\ref{fig:gevall}, both because of the miscounting problem and the lower sensitivity of GOES as compared with RHESSI during quiet times.
This suggests that a more comprehensive study of the RHESSI hard X-ray flare statistics, 
including new data from cycle~24, will be interesting.

\section{Conclusions}

The NOAA database, as checked by the RHESSI data after 2002, shows long-term variability of the flare productivity.
Specifically, active regions in 2004-2005 had flare productivities (as we define them) about twice as large as those at other times.
The most flare-productive regions immediately prior to this epoch were significantly less productive in general, with no regions in the first half of cycle~23 producing more than 10 M or X-class flares.
This limit is strikingly smaller than that of prior or subsequent intervals.
These effects do not appear to depend repeatably  on solar-cycle phase, since the 2004-2005 increase was unique in the time interval since 1975.
Wheatland (2001) had noted the existence of significant variations in the occurrence rates in individual active regions, and the pattern we have found seems consistent with that.
We do not have any speculations as to the origin of this effect, but its timing just prior to the ``anomalous'' cycle minimum between cycles~23 and~24 suggests a possible relationship.
On the plausible idea that flare occurrence results from twists imposed on the solar magnetic field prior to eruption, this effect might provide a clue to the dynamo action in the solar interior.
The lack of a solar-cycle dependence in the flare productivity of active regions deserves mention. 
Except for the surge in 2004-2005, the number of (M, X)-class flares, in regions that produce them, appears to remain in the range 1-3 flares per region at all other times.
This is consistent with the idea that the ability of an active region to produce an energetic flare is intrinsic to its own structure, rather than to interactions with other structures.

Databases now being produced, such as that from RHESSI and the ``Latest Events'' catalog
of S.~Freeland (\url{http://www.lmsal.com/solarsoft/last_events/}) offer much superior metadata, and it would be possible to extend the latter to cover global event occurrence by use of positions from the STEREO ({\it Solar Terrestrial Relations Observatories}) data (\textit{e.g.}, Kaiser {\it et al.}, 2008).
\nocite{2008SSRv..136....5K}
This would help to clarify the heliographic biases present in event identifications inevitably present in any Earth-based observational material.
We suggest revisiting this question in about 2020, when the statistics for cycle 24 will be complete.
\nocite{2002SoPh..210....3L}

 \begin{acks}
Authors Hudson and McTiernan acknowledge support from NASA under Contract NAS5-98033 for RHESSI.
Author Fletcher was supported by STFC rolling grant ST/I001808/1 and by the EC-funded FP7 project HESPE (FP7- 2010-SPACE-1-263086).
\end{acks}


%

%

%

%

%
%

%
%

 \bibliographystyle{spr-mp-sola}

\bibliography{cycle}  
%
%
%
%

\end{article} 
\end{document}